\pdfoutput=1
\documentclass[11pt]{article}
\usepackage{amssymb,latexsym,amsmath}
\usepackage{graphicx}
\usepackage{cite}
\newtheorem{theo}{Theorem}

\newtheorem{lemm}[theo]{Lemma}

\newtheorem{prop}[theo]{Proposition}
\def\ii{\mathrm{i}} 
\def\ee{\mathrm{e}} 
\newcommand{\myatop}[2]{\genfrac{}{}{0pt}{}{#1}{#2}}
\def\mybox{\hfill$\Box$}%
\begin{document}
\begin{center}
{\Large \bf
A Wigner distribution function for finite oscillator systems}\\[5mm]
{\bf J.\ Van der Jeugt} \\[1mm]
Department of Applied Mathematics and Computer Science,
Ghent University,\\
Krijgslaan 281-S9, B-9000 Gent, Belgium\\[1mm]
E-mail: Joris.VanderJeugt@UGent.be
\end{center}

\vskip 10mm
\noindent
Short title: Wigner function for finite systems



\begin{abstract}
We define a Wigner distribution function for a one-dimensional finite quantum system, in which 
the position and momentum operators have a finite (multiplicity-free) spectrum.
The distribution function is thus defined on discrete phase-space, i.e.\ on a finite discrete square grid.
These discrete Wigner functions possess a number of properties similar to the Wigner function
for a continuous quantum system such as the quantum harmonic oscillator.
As an example, we consider the so-called $\mathfrak{su}(2)$ oscillator model in dimension $2j+1$, which is
known to tend to the canonical oscillator when $j$ tends to infinity.
In particular, we compare plots of our discrete Wigner functions for the $\mathfrak{su}(2)$ oscillator with
the well known plots of Wigner functions for the canonical quantum oscillator.
This comparison supports our approach to discrete Wigner functions.
\end{abstract}

\section{Introduction}

The Wigner distribution function is a quasiprobability distribution introduced by Wigner~\cite{Wigner1932} in order to study
quantum corrections to classical statistical mechanics. 
It can be considered as a phase-space formulation of quantum mechanics, and has since been investigated and applied extensively.
The Wigner function allows to compute expectation values of quantum mechanical observables in a classical way (as an integral)
rather than through the operator formalism.

The amount of literature on Wigner functions is huge. Some of the classical papers dealing with the fundamentals of the 
Wigner representation of quantum mechanics are Refs.~\cite{Cohen1966,Tatarskii1983,Hillery1984,Lee1995}; see also~\cite{Curtright2012}
for a historical account. 

For a (one dimensional) quantum mechanical system described by a continuous wavefunction $\psi(q)$, where $q$ is the position and the momentum
is denoted by $p$, the Wigner distribution function is usually defined as
\begin{equation}
W(p,q)= W_\psi (p,q) = \frac{1}{2\pi} \int_{-\infty}^{\infty} \psi^*(q+\frac{y}{2})\psi(q-\frac{y}{2})\ee^{\ii p y} dy
\label{Wdef-cont}
\end{equation}
(in the convention $\hbar=1$).
Moyal~\cite{Moyal1949} emphasized how the Wigner distribution function provides the integration measure in phase-space, to yield 
expectation values from phase-space functions $G(p,q)$ (as a classical integral with the Wigner function as probability density
function) for the corresponding quantum mechanical expectation value of the suitable ordered operator expression ${\hat G}({\hat p},{\hat q})$.
In other words:
\begin{equation}
\langle \psi | {\hat G}({\hat p},{\hat q}) | \psi \rangle = \int\int W(p,q) G(p,q) dp dq.
\label{Wdef-integral}
\end{equation}
The correspondence (or association rule) between $G(p,q)$ and ${\hat G}({\hat p},{\hat q})$ is usually that of Weyl, although there are
many other possibilities (leading to other types of distribution functions, see~\cite{Lee1995}).
In~\cite{Tatarskii1983}, one finds a clear analysis of how the condition~\eqref{Wdef-integral} for analytic functions $G$ of $p$ and $q$
leads to the integral form~\eqref{Wdef-cont}, provided Weyl's association rule is followed, and provided one uses the canonical
commutation relation $[{\hat q}, {\hat p}]=\ii$.

In the present paper we will define a Wigner distribution function for a one-dimensional quantum system in a {\em finite} Hilbert space.
In that situation, the position and momentum operators ${\hat q}$ and ${\hat p}$ have a finite (discrete) spectrum, say $q_0,q_1,\ldots,q_N$ 
and $p_0,p_1,\ldots,p_N$ respectively. 
The position wavefunction $\phi_n(q)$ is then a discrete wavefunction with finite support $q \in \{ q_0,q_1,\ldots,q_N \}$ (and similarly
for the momentum wavefunctions).
In the finite case, the canonical commutation relation can no longer hold.

The question is now how to define the analogue of a Wigner function for this finite discrete case.
One approach could be through a discrete version of~\eqref{Wdef-cont}, replacing the integral by a sum and the exponential function by
the kernel of a discrete Fourier transform related to the finite quantum system. It is, however, not clear how to deal with the 
shifts in the arguments of the functions.
Our own approach starts from the analogue of~\eqref{Wdef-integral}, and requires that for stationary states $|n\rangle$ of the system, one has
\begin{equation}
\langle n | {\hat G}({\hat p},{\hat q}) | n \rangle = \sum_{k=0}^N\sum_{l=0}^N W(n;p_k,q_l) G(p_k,q_l) 
\label{Wdef-sum}
\end{equation}
for an appropriate class of functions $G(p,q)$. 
We will show in section~2 that this leads to a unique definition for the discrete Wigner functions.
Moreover, these Wigner functions satisfy a number of nice properties, analogous to continuous Wigner functions: they are real, they are
normalized over the discrete support in $(p,q)$ phase-space, and their marginals yield position and momentum probability distributions.

Of course, we are not the first to investigate the possibility of extending Wigner functions to finite-state quantum mechanics.
One approach is given by Wootters~\cite{Wootters1987}. In his approach, the Wigner function satisfies a number of characteristic features.
However, the technique works only in the case that the axis observables (say, in this case $p$ and $q$) are mutually unbiased,
which is a rather strong requirement.
In other approaches~\cite{Barker2000,Ozaktas2000}, the support of position (or momentum) operators is often forced to the set
$\{0,1,\ldots,N\}$, the Hamiltonian operator is replaced by an appropriate difference operator, and one uses the common (fractional) discrete
Fourier transform in order to define a Wigner function.
Also in the review paper~\cite{Vourdas2004}, the Wigner function corresponding to an operator is the discrete Fourier transform
of the matrix elements of that operator.
A group theoretical approach was given in~\cite{Atak1998}: here, the Wigner operator is defined by extending~\eqref{Wdef-cont} to
an integral over a Lie group (using the invariant measure). The finite systems then arise through finite-dimensional representations
of the Lie group.
Clearly, all of these approaches (and their generalizations) are valuable, each of them having certain advantages and being subject
to certain constraints but also showing some shortcomings compared to the continuous case.

We believe that our approach, presented here in section~2, is new. 
In a way, one could say that our approach is algebraic rather than geometric.
In various articles where the Wigner framework is addressed for finite Hilbert spaces, one tries to associate to the operator ${\hat p}$ 
a generator of (finite) translations in $q$ in the phase space $(p,q)$ (and similar for ${\hat q}$). 
In such a geometric setting one is forced to add boundary conditions: 
usually periodic boundary conditions leading to the finite phase space becoming a torus~\cite{Hannay,Galetti,Bouzouina,Rivas}.
Translations are then replaced by a unitary quantum propagation matrix, and the Wigner function is a transform of the corresponding unitary operator.
This theory is well developed, but the notion of position and momentum operators on a finite torus remains elusive~\cite{Rivas}.

Our algebraic approach is different, and assumes in particular that the (self-adjoint) position and momentum operators ${\hat q}$ and ${\hat p}$ are
elements of some algebra, and that the spectrum of ${\hat q}$ and ${\hat p}$ in finite-dimensional unitary representations of this algebra
is explicitly known and non-degenerate.
Once this is the case, our definition of a Wigner function is easy to apply.
The obtained Wigner function is real, satisfies the common normalization property, and the 
$q$ and $p$ probability distributions are given by marginals of the Wigner function.
In order to have reflection symmetries, one should however assume further requirements for the finite quantum system.
Also, there is no straightforward analogue of Galilei-covariance.

In the following section, we motivate and describe our approach. 
The main condition is~\eqref{Wab}, and it is shown how this determines the Wigner function unambiguously.
The essential properties satisfied by the Wigner function are shown in Proposition~2.
The conditions for phase-space symmetries of the Wigner function are described in Proposition~3.
Section~3 is dealing with an example: the $\mathfrak{su}(2)$ oscillator model, which is the simplest example of a finite oscillator
(and whose algebraic structure coincides with the angular momentum algebra).
Here, in the representation of dimension $2j+1$, the position and momentum operators have the spectrum $\{-j,-j+1,\ldots,+j\}$,
and the position (momentum) wavefunctions are given in terms of symmetric Krawtchouk polynomials.
We illustrate how to compute the Wigner function, and present some plots of our discrete Wigner functions, 
comparing them to the usual Wigner functions for the canonical quantum oscillator.
This example clearly supports the significance of our approach.

\section{The general setting}

Consider a one-dimensional quantum system described by a Hamiltonian operator ${\hat H}$, and let the position and momentum operator be denoted by ${\hat q}$ and ${\hat p}$.
We shall assume that ${\hat H}$, ${\hat q}$ and ${\hat p}$ are self-adjoint elements of some algebra (with a $\star$-operation), and that the unitary representations of this algebra are finite-dimensional, so that we are dealing with a finite quantum system.
In the case that these operators satisfy the Hamilton-Lie equations
\begin{equation}
[\hat H, \hat q] = -\ii \hat p, \qquad [\hat H,\hat p] = \ii \hat q,
\label{Hqp}
\end{equation}
and when ${\hat H}$ has an equidistant spectrum, the quantum system is referred to as a finite quantum 
oscillator~\cite{Atak2001,Atak2005,JSV2011,JV2012}.
In the current section, however, we need not assume this last condition.

Let us consider a representation space $V$ of dimension $N+1$, and denote the (normalized) eigenvectors of ${\hat H}$ by $|n\rangle$:
\begin{equation}
{\hat H} |n\rangle = E_n |n\rangle, \qquad (n=0,1,\ldots,N).
\end{equation}
$V$ is a Hilbert space, and 
\begin{equation}
\langle n' | n\rangle = \delta_{n',n}.
\end{equation}
These eigenvectors are referred to as the stationary states.
In the basis $|n\rangle$ ($n=0,1,\ldots,N$), the operators ${\hat q}$ and ${\hat p}$ are represented by Hermitian matrices, whose eigenvalues 
are real: these eigenvalues correspond to the finite spectrum of these operators, or the ``possible position and momentum values''.
Let us denote the eigenvalues of ${\hat q}$ by $q_k$ ($k=0,1,\ldots,N$), and the corresponding (orthonormal) eigenvectors by $|q_k\rangle$. We shall assume that all eigenvalues are different and simple (i.e.\ non-degenerate spectra).
So we have the following situation:
\begin{equation}
{\hat q} |q_k\rangle = q_k |q_k\rangle, \qquad (k=0,1,\ldots,N).
\end{equation}
The expansion of the ${\hat q}$ eigenstates in the basis $|n\rangle$ ($n=0,1,\ldots,N$) is denoted by
\begin{equation}
|q_k\rangle = \sum_{n=0}^N \phi_n(q_k) |n\rangle, \qquad (k=0,1,\ldots,N).
\label{qk}
\end{equation}
The orthogonality of the vectors $|q_k\rangle$ implies
\begin{equation}
\sum_{n=0}^N \phi_n^*(q_l) \phi_n(q_k)=\delta_{k,l}, \qquad (k,l=0,1,\ldots,N),
\label{qorth1}
\end{equation}
so the numbers $\phi_n(q_k)$ form a unitary matrix, and thus they also satisfy
\begin{equation}
\sum_{k=0}^N \phi_n(q_k) \phi_{n'}^*(q_k)=\delta_{n,n'}, \qquad (n,n'=0,1,\ldots,N).
\label{qorth2}
\end{equation}
The relations inverse to~\eqref{qk} read
\begin{equation}
|n\rangle = \sum_{k=0}^N \phi_n^*(q_k) |q_k\rangle, \qquad (n=0,1,\ldots,N).
\label{nq}
\end{equation}
The discrete function $\phi_n(q)$, defined over $q\in\{q_0,q_1,\ldots,q_N\}$, can be interpreted as the position wavefunction when the system is in the $n$th stationary state.

Completely similar, we shall assume that the eigenvalues of ${\hat p}$ are given by the mutually distinct values $p_k$ ($k=0,1,\ldots,N$), and denote the corresponding (orthonormal) eigenvectors by $|p_k\rangle$. 
So 
\begin{equation}
{\hat p} |p_k\rangle = p_k |p_k\rangle, \qquad (k=0,1,\ldots,N), 
\end{equation}
and we denote
\begin{equation}
|p_k\rangle = \sum_{n=0}^N \psi_n(p_k) |n\rangle, \qquad (k=0,1,\ldots,N).
\label{pk}
\end{equation}
The coefficients $\psi_n(p_k)$ satisfy the same orthogonality relations as the $\phi_n(q_k)$.

\vskip 2mm
Before describing our construction of a Wigner distribution function, it is useful to summarize the assumptions (necessary to perform the construction) in an unambiguous way:
\begin{enumerate}
\item ${\hat H}$, ${\hat q}$ and ${\hat p}$ are self-adjoint elements of some $\star$-algebra.
\item The $\star$-algebra has a class of finite-dimensional unitary representations.
\item For such a representation of dimension $N+1$ (with representation space $V$), a basis of eigenvectors of ${\hat H}$ is known and denoted by $|n\rangle$
($n=0,1,\ldots,N$). 
\item The eigenvalues $q_0, q_1, \ldots, q_N$ (resp.\ $p_0, p_1, \ldots, p_N$) of ${\hat q}$ (resp.\ ${\hat p}$) in each representation $V$ are known explicitly. 
\end{enumerate}
The remaining assumption
\begin{enumerate}
\setcounter{enumi}{4}
\item The operators ${\hat H}$, ${\hat q}$ and ${\hat p}$ satisfy~\eqref{Hqp} and the spectrum of ${\hat H}$ in $V$ is equidistant.
\end{enumerate}
is not required, but when it is satisfied one usually speaks of a finite quantum oscillator.
Note that the explicit knowledge of the discrete wavefunctions $\phi_n(q)$ or $\psi_n(p)$ is not required.
On the other hand, ${\hat q}$ and ${\hat p}$ are operators, so their action in representations (in particular on basis elements
$|n\rangle$) is obviously assumed to be known.

\vskip 2mm
The purpose is now to construct a Wigner distribution function corresponding to the state $|n\rangle$.
For a continuous quantum system in a stationary state $\Phi_n$, the Wigner function $W_n(p,q)$ is defined in the $(p,q)$-phase space in such a way that averages of quantum ``observables'' (i.e.\ functions of ${\hat q}$ and ${\hat p}$) correspond to averages of their classical counterparts, i.e.\
\begin{equation}
\langle \Phi_n | {\hat G}({\hat p},{\hat q}) | \Phi_n\rangle =
\int\int W_n(p,q) G(p,q) dpdq,
\end{equation}
for a class of functions $G(p,q)$.
So in the discrete case, it is natural to require that the Wigner function $W(n;p_k,q_l)$ should (for every $n$) be a function of the discrete values $(p_k,q_l)$ such that
\begin{equation}
\langle n | {\hat G}({\hat p},{\hat q}) | n\rangle = \sum_{k=0}^N \sum_{l=0}^N W(n;p_k,q_l) G(p_k,q_l).
\label{W1}
\end{equation}
The correspondence between ordinary functions $G(p,q)$ and operator functions ${\hat G}({\hat p},{\hat q})$ will be described soon, and is just the usual Weyl correspondence.
Sometimes, the notion of Wigner function is extended to the ``cross Wigner distribution function''~\cite{Ozaktas2000}, 
which in our case would be defined through
\begin{equation}
\langle n' | {\hat G}({\hat p},{\hat q}) | n\rangle = \sum_{k=0}^N \sum_{l=0}^N W(n',n;p_k,q_l) G(p_k,q_l).
\label{W2}
\end{equation}
Because of the finite setting, one can represent the (cross) Wigner distribution function as an $(N+1)\times(N+1)$ matrix ${\mathbf W}(n)_{0\leq k,l \leq N}$ (or ${\mathbf W}(n',n)_{0\leq k,l \leq N}$) with matrix elements
\begin{equation}
{\mathbf W}(n)_{k,l}= W(n;p_k,q_l) \qquad (\hbox{or } {\mathbf W}(n',n)_{k,l}= W(n',n;p_k,q_l)).
\end{equation}

We want to show that the condition~\eqref{W1} is in a certain way sufficient to fix the Wigner function $W(n;p_k,q_l)$ uniquely. For this purpose, let us first recall the Weyl correspondence between ordinary functions $G(p,q)$ and operator functions ${\hat G}({\hat p},{\hat q})$.
This is commonly defined in such a way~\cite{Weyl1927,Tatarskii1983} that there is a correspondence between
\begin{equation}
\ee^{\lambda p + \mu q}  \leftrightarrow \ee^{\lambda \hat p + \mu \hat q}.
\end{equation}
So for a monomial function, denoted by
\[
G_{a,b}(p,q)=p^a q^b,\qquad (a,b=0,1,\ldots)
\]
the corresponding operator function would be given by
\begin{equation}
{\hat G}_{a,b}({\hat p},{\hat q}) = \frac{1}{\binom{a+b}{a}} 
 \left. (\lambda \hat p + \mu \hat q)^{a+b} \right|_{\lambda^a\mu^b},
\end{equation}
where the last action stands for taking the coefficient of $\lambda^a\mu^b$ in the expansion of $(\lambda \hat p + \mu \hat q)^{a+b}$. For example, for $G_{2,2}(p,q)=p^2q^2$ one finds
\begin{equation}
{\hat G}_{2,2}({\hat p},{\hat q}) =\frac{1}{6}(\hat p \hat p \hat q \hat q + \hat p \hat q \hat p \hat q + \hat p \hat q \hat q \hat p + \hat q \hat p \hat p \hat q + \hat q \hat p \hat q \hat p + \hat q \hat q \hat p \hat p).
\end{equation}

In the current case we are dealing with finite representation matrices. This implies, for instance, that the representation matrix of $\hat q$, with eigenvalues $q_0,q_1,\ldots,q_N$, satisfies the Caley-Hamilton relation, which could be written as
\[
({\hat q} -q_0)({\hat q} -q_1)\cdots ({\hat q} -q_N)=0,
\]
in other words ${\hat q}^{N+1}$ (and as a consequence also higher powers of $\hat q$) is expressible into powers of $\hat q$ up to $N$. 
For this reason, it seems to be acceptable that the ``quantum observables'' are taken to be (linear combinations of) functions $G_{a,b}$ with
$0\leq a,b\leq N$. 
In other words, we require that 
\begin{equation}
\langle n | {\hat G}_{a,b}({\hat p},{\hat q}) | n\rangle = \sum_{k=0}^N \sum_{l=0}^N W(n;p_k,q_l) G_{a,b}(p_k,q_l)
\quad\hbox{ for }\quad 0\leq a,b \leq N.
\label{Wab}
\end{equation}
Equation~\eqref{Wab} will determine the Wigner function in an unambiguous way.
For this purpose, let us now define the $(N+1)\times(N+1)$ matrix ${\mathbf Z}(n)_{0\leq a,b\leq N}$ by
\begin{equation}
{\mathbf Z}(n)_{a,b} = \langle n | {\hat G}_{a,b}({\hat p},{\hat q}) |n \rangle, \qquad (a,b=0,1,\ldots,N).
\label{defZ}
\end{equation}
From~\eqref{Wab} it follows that 
\begin{equation}
{\mathbf Z}(n)_{a,b} = \sum_{k=0}^N \sum_{l=0}^N {\mathbf W}(n)_{k,l} p_k^a q_l^b
\label{ZW1}
\end{equation}
should hold for all $a,b=0,1,\ldots,N$. This implies that the matrix ${\mathbf Z}(n)$ can be written as the matrix product
\begin{equation}
{\mathbf Z}(n)= {\mathbf V}(p_0,p_1\ldots,p_N)^T {\mathbf W}(n) {\mathbf V}(q_0,q_1\ldots,q_N),
\label{ZW2}
\end{equation}
where ${\mathbf V}(x_0,x_1\ldots,x_N)$ stands for the $(N+1)\times(N+1)$ Vandermonde matrix corresponding to the values $x_0,x_1\ldots,x_N$, i.e.\ ${\mathbf V}(x_0,x_1\ldots,x_N)_{r,s} = x_r^s$, or
\begin{equation}
{\mathbf V}(x_0,x_1\ldots,x_N) = \left[
 \begin{array}{cccc}
 1 & x_0^1 & \cdots & x_0^N \\
 1 & x_1^1 & \cdots & x_1^N \\
 \vdots & \vdots &  & \vdots \\
 1 & x_N^1 & \cdots & x_N^N 
 \end{array} \right].
\end{equation}
Since Vandermonde matrices (on mutually distinct values $x_r$) are invertible, we can write
\begin{equation}
{\mathbf W}(n) = {\mathbf V}(p_0,p_1\ldots,p_N)^{-T}{\mathbf Z}(n){\mathbf V}(q_0,q_1\ldots,q_N)^{-1},
\label{main}
\end{equation}
where, as usual, ${\mathbf A}^{-T}$ stands for the inverse of the transpose of a matrix ${\mathbf A}$.

Equation~\eqref{main} gives a method to compute the Wigner distribution function $W(n;p_k,q_l)$, or equivalently the matrix elements of the Wigner matrix ${\mathbf W}(n)$. Indeed, the matrix ${\mathbf Z}(n)$ can in principle be computed by means of its definition~\eqref{defZ}. Then it is a matter of computing Vandermonde matrices, their inverses, and performing matrix multiplications. In the following section we shall discuss an example to illustrate this procedure.
Here, we shall first determine a number of properties of the defined Wigner distribution function.

Before that, note that there exists an explicit formula for the inverse of a Vandermonde matrix.
\begin{lemm}
Let ${\mathbf V} = (x_i^k)_{0\leq i,k \leq N}$, then its inverse ${\mathbf V}^{-1}$ has matrix elements
\[
{\mathbf V}^{-1}_{k,j} = (-1)^{N-k} e_{N-k}^{(j)}\; /\prod_{l=0\, (l\ne j)}^N (x_j-x_l),
\qquad (k,j=0,1,\ldots,N).
\]
Herein, $e_r^{(j)}$ is the $r$th elementary function in the variables $x_0,\ldots,x_{j-1},x_{j+1},\ldots,x_N$ (omitting $x_j$).
\end{lemm}

\noindent {\bf Proof.} 
The proof is similar to that of Property (3.6) in Chapter~I of Macdonald~\cite{Macdonald}.
Let ${\mathbf V} = (x_i^k)_{0\leq i,k \leq N}$, 
\begin{equation}
{\mathbf M}= \left( (-1)^{N-k} e_{N-k}^{(j)}\right)_{0\leq k,j\leq N},
\end{equation}
and ${\mathbf D}$ the diagonal matrix with elements ${\mathbf D}_{ii}= \prod_{l=0\, (l\ne i)}^N (x_i-x_l)$.
Then the statement of the Lemma is equivalent to saying that
\[
{\mathbf V} {\mathbf M}= {\mathbf D}.
\]
To prove the last assertion, observe that
\begin{equation}
({\mathbf V} {\mathbf M})_{ij} = \sum_{k=0}^N x_i^k (-1)^{N-k} e_{N-k}^{(j)}
= \sum_{k=0}^N x_i^{N-k} (-1)^{k} e_{k}^{(j)} = x_i^N \sum_{k=0}^N e_{k}^{(j)} \left( -\frac{1}{x_i}\right)^{k} .
\end{equation}
But according to the generating function for elementary symmetric functions~\cite[I~(2.2)]{Macdonald}, this can be written as
\begin{equation}
({\mathbf V} {\mathbf M})_{ij} = x_i^N \prod_{l=0\, (l\ne j) }^N \left( 1+ x_l\left( -\frac{1}{x_i}\right) \right) 
= \prod_{l=0\, (l\ne j) }^N (x_i-x_l),
\end{equation}
and thus the assertion follows.
\mybox

We shall now establish a number of properties of the Wigner distribution function $W(n;p,q)$.
\begin{prop} {\ }
\begin{enumerate}
\item $W(n;p,q)$ is real on the support $\{p_0,\ldots,p_N\}\times \{q_0,\ldots,q_N\}$.
\item The position and momentum probability distributions are given by marginals:
\begin{align}
& \sum_{k=0}^N W(n;p_k,q_l) = |\phi_n(q_l)|^2, \qquad (l=0,1,\ldots,N),  \label{q-marginal}\\
& \sum_{l=0}^N W(n;p_k,q_l) = |\psi_n(p_k)|^2, \qquad (k=0,1,\ldots,N) . \label{p-marginal}
\end{align}
\item The distribution function satisfies
\begin{equation}
\sum_{k=0}^N \sum_{l=0}^N W(n;p_k,q_l)=1,
\label{Wis1}
\end{equation}
\end{enumerate}
\end{prop}

\noindent {\bf Proof.} 
Since ${\hat p}^\dagger=\hat p$, ${\hat q}^\dagger=\hat q$, it follows that ${\hat G}_{a,b}(\hat p, \hat q)^\dagger = 
{\hat G}_{a,b}(\hat p, \hat q)$. So the matrix ${\mathbf Z}(n)$ in~\eqref{defZ} is real; since the Vandermonde matrices are real, it follows from~\eqref{main} that the matrix ${\mathbf W}(n)$ is real, so the first statement follows.

To prove~\eqref{q-marginal}, first note that if for numbers $A_l$, $B_l$ ($l=0,1,\ldots,N$)
\begin{equation}
\sum_{l=0}^N A_l q_l^b = \sum_{l=0}^N B_l q_l^b \quad\hbox{ for all } \quad b=0,1,\ldots,N,  \label{AB}
\end{equation}
then this implies that $A_l=B_l$ for all $l=0,1,\ldots,N$. Indeed, if we denote by ${\mathbf A}$ the column matrix with elements $A_0, A_1,\ldots, A_N$ and by ${\mathbf B}$ the column matrix with elements $B_0, B_1,\ldots, B_N$, then~\eqref{AB} is written as 
\[
{\mathbf V}(q_0,q_1\ldots,q_N)^T {\mathbf A} = {\mathbf V}(q_0,q_1\ldots,q_N)^T {\mathbf B},
\]
and since the Vandermonde matrices are invertible, the conclusion follows.
Consider now the matrix element of ${\hat G}_{0,b}(\hat p, \hat q)={\hat q}^b$, and expand the basis vectors $|n\rangle$ according to~\eqref{nq}:
\begin{align}
\langle n | {\hat G}_{0,b}(\hat p, \hat q) | n \rangle & =
\langle n |  {\hat q}^b | n \rangle =
\sum_{k=0}^N \sum_{l=0}^N \langle q_l | \phi_n(q_l) {\hat q}^b \phi_n^*(q_k) | q_k\rangle \nonumber \\
& = \sum_{k=0}^N \sum_{l=0}^N \phi_n(q_l)\phi_n^*(q_k) \langle q_l |  {\hat q}^b | q_k\rangle
=  \sum_{k=0}^N \sum_{l=0}^N \phi_n(q_l)\phi_n^*(q_k) q_k^b\langle q_l | q_k\rangle \nonumber \\
&= \sum_{l=0}^N  |\phi_n(q_l)|^2 q_l^b.
\end{align}
From~\eqref{Wab} we have, with $G_{0,b}(p,q)=q^b$,
\[
\langle n | {\hat G}_{0,b}(\hat p, \hat q) | n \rangle  = \sum_{k=0}^N \sum_{l=0}^N
W(n;p_k,q_l) q_l^b,
\]
so
\[
\sum_{l=0}^N  |\phi_n(q_l)|^2 q_l^b = \sum_{l=0}^N \left(\sum_{k=0}^N W(n;p_k,q_l)\right) q_l^b
\]
for all $b=0,1,\ldots,N$. This is of the form~\eqref{AB}, so we can conclude $\sum_{k=0}^N W(n;p_k,q_l) = |\phi_n(q_l)|^2$
for all $l=0,1,\ldots,N$, proving~\eqref{q-marginal}. The proof of~\eqref{p-marginal} is analogous. 
Finally, summing over all $l$-values in~\eqref{q-marginal} and using~\eqref{qorth2} yields the third assertion.
\mybox

The above results can be extended to the cross Wigner function:
\begin{equation}
\sum_{k=0}^N W(n',n;p_k,q_l) = \phi_{n'}(q_l)\phi_n^*(q_l), \qquad 
\sum_{l=0}^N W(n',n;p_k,q_l) = \psi_{n'}(p_k)\psi_n^*(p_k), 
\end{equation}
and thus
\begin{equation}
\sum_{k=0}^N\sum_{l=0}^N W(n',n;p_k,q_l) =\delta_{n',n}.
\end{equation}

Before discussing certain symmetry properties of the Wigner function, let us extend~\eqref{AB}.
If 
\begin{equation}
\sum_{k=0}^N \sum_{l=0}^N A_{kl} p_k^a q_l^b = \sum_{k=0}^N \sum_{l=0}^N B_{kl} q_k^a q_l^b \quad\hbox{ for all } \quad a,b=0,1,\ldots,N,  \label{ABext}
\end{equation}
then this implies that $A_{kl}=B_{kl}$ for all $k,l=0,1,\ldots,N$. The argument is the same as in~\eqref{AB}
(or follows matrix multiplications like~\eqref{ZW1}-\eqref{ZW2}). 

\begin{prop}
Assume that the spectra of $\hat p$ and $\hat q$ are symmetric in space, i.e.\
\begin{equation}
p_k = -p_{N-k}, \qquad q_k=-q_{N-k} \qquad (k=0,1,\ldots,N).
\label{sym}
\end{equation}
Moreover, assume that the matrix representation of $\hat p$ and $\hat q$ in the basis $|n\rangle$ is tridiagonal with zeros on the diagonal, i.e.\
\begin{equation}
\langle n' | \hat p |n\rangle \hbox{ and } \langle n' | \hat q |n\rangle \hbox{ nonzero only for } n'=n\pm 1.
\label{tridia}
\end{equation}
Then the Wigner function satisfies the following symmetries:
\begin{equation}
W(n;p_k,q_l) = W(n;p_{N-k},q_l) = W(n;p_k,q_{N-l}) = W(n;p_{N-k},q_{N-l}).
\label{Wsym}
\end{equation}
\end{prop}
The first assumption~\eqref{sym} looks rather natural.
At first sight, the second assumption~\eqref{tridia} looks strange, but in fact for many known finite oscillator models this is satisfied. Let us now proceed with the proof of this Proposition.\\
\noindent {\bf Proof.} 
If for an operator $\hat x$ (or its matrix representation) one has $\langle n' | \hat x |n\rangle$ is nonzero only for $n'=n\pm 1$, then it follows by matrix multiplications that for its multiples
\begin{align}
&\langle n' | {\hat x}^{2r} |n\rangle=0 \hbox{ when } n'+n \hbox{ is odd},\\
&\langle n' | {\hat x}^{2r+1} |n\rangle=0 \hbox{ when } n'+n \hbox{ is even}.
\end{align}
From this argument and~\eqref{defZ} one can deduce that 
\begin{equation}
{\mathbf Z}(n)_{a,b} =0 \hbox{ when $a$ is odd or when $b$ is odd}.
\label{odd}
\end{equation}
Using~\eqref{sym} in~\eqref{ZW1}, one finds
\[
{\mathbf Z}(n)_{a,b} = \sum_{k=0}^N \sum_{l=0}^N {\mathbf W}(n)_{k,l} (-p_{N-k})^a q_l^b =
(-1)^a \sum_{k=0}^N \sum_{l=0}^N {\mathbf W}(n)_{N-k,l} p_k^a q_l^b.
\]
So if $a$ is even, we have
\begin{equation}
\sum_{k=0}^N \sum_{l=0}^N {\mathbf W}(n)_{k,l} p_k^a q_l^b = \sum_{k=0}^N \sum_{l=0}^N {\mathbf W}(n)_{N-k,l} p_k^a q_l^b,
\label{WW}
\end{equation}
and if $a$ is odd, this also holds because both sides are 0 by~\eqref{odd}.
Thus~\eqref{WW} is valid for all $a=0,1,\ldots,N$ (and for all $b=0,1,\ldots,N$), so by~\eqref{ABext}:
\[
{\mathbf W}(n)_{k,l} = {\mathbf W}(n)_{N-k,l}
\]
for all $k,l=0,1,\ldots,N$, which is the first statement in~\eqref{Wsym}. The others are proved in a similar way. \mybox

So far, we have mainly concentrated on Wigner functions for pure states $|n\rangle$. For a superposition
\[
|\Phi\rangle= \sum_{n=0}^N c_n |n\rangle,
\]
which we can assume to be orthonormal, the Wigner function  is required to satisfy
\begin{equation}
\langle \Phi | {\hat G}_{a,b}({\hat p},{\hat q}) | \Phi \rangle = \sum_{k=0}^N \sum_{l=0}^N W(\Phi;p_k,q_l) G_{a,b}(p_k,q_l)
\quad\hbox{ for }\quad 0\leq a,b \leq N.
\label{WPhi}
\end{equation}
Expansion of the left hand side gives:
\begin{align}
& \sum_{n'=0}^N \sum_{n=0}^N c_{n'}^* c_n \langle n' | {\hat G}_{a,b}({\hat p},{\hat q}) | n \rangle \nonumber \\
&= \sum_{n'=0}^N \sum_{n=0}^N c_{n'}^* c_n \sum_{k=0}^N\sum_{l=0}^N W(n',n;p_k,q_l) G_{a,b}(p_k,q_l) \nonumber \\
&= \sum_{k=0}^N\sum_{l=0}^N \left( \sum_{n'=0}^N \sum_{n=0}^N c_{n'}^* c_n W(n',n;p_k,q_l) \right) p_k^a q_l^b
\end{align}
Since this must be equal to the right hand side of~\eqref{WPhi}, it follows from~\eqref{ABext} that
\begin{equation}
W(\Phi;p_k,q_l) = \sum_{n'=0}^N \sum_{n=0}^N c_{n'}^* c_n W(n',n;p_k,q_l),
\end{equation}
so here the cross Wigner functions play a role.

\section{Example: the $\mathfrak{su}(2)$ oscillator model}

Inspired by the need of a finite quantum oscillator in quantum optics, Atakishiyev {\em et al}~\cite{Atak2001,Atak2005} 
developed the $\mathfrak{su}(2)$ oscillator model, which can be considered as the simplest example of a finite oscillator.
In terms of the standard $\mathfrak{su}(2)$ basis $J_0, J_+, J_-$ (with commutation relations $[J_0,J_\pm]=\pm J_\pm$, $[J_+,J_-]=2J_0$), and working in the representation space $V=V_j$ labeled by a nonnegative integer of half-integer $j$ ($2j\in{\mathbb Z}_+$),
the Hamiltonian, position and momentum operators are defined by
\[
{\hat H}= J_0+j+\frac12,\qquad {\hat q}=\frac12 (J_++J_-), \qquad {\hat p}=\frac{\ii}{2} (J_+-J_-),
\] 
satisfying~\eqref{Hqp}.
The basis states of $V_j$ in the ``angular momentum'' notation are $|j,m\rangle$ ($m=-j,-j+1,\ldots,+j$), with the well known action
\[
J_0 |j,m\rangle = m |j,m\rangle, \qquad J_\pm |j,m\rangle = \sqrt{(j\mp m)(j\pm m+1)} |j,m\pm1\rangle.
\]
Thus also the matrices of ${\hat p}$ and ${\hat q}$ in this basis are clear from this action.
Following the notation of the previous section, we have $N=2j$, and the Hamiltonian eigenstates are denoted by
\begin{equation}
|n\rangle = |j,m\rangle \hbox{ with } n=j+m \qquad (n=0,1,\ldots,N=2j).
\end{equation}

The eigenvalues and eigenvectors of ${\hat q}$ (and ${\hat p}$) have been determined in~\cite{Atak2001}. One has
\[
q_k = -j+k \qquad (k=0,1,\ldots,N),
\]
with (following the notation of~\eqref{qk}) 
\begin{equation}
\phi_n(q) = \frac{(-1)^n}{2^j} \sqrt{\binom{2j}{n} \binom{2j}{j+q}} K_n(j+q;\frac12,2j).
\end{equation}
Herein, $K_n$ is the Krawtchouk polynomial~\cite{Koekoek,Ismail,Andrews}:
\begin{equation}
K_n(x;p,N) = {\;}_2F_1 \left( \myatop{-n,-x}{-N} ; \frac{1}{p} \right).
\end{equation}
So the discrete position wavefunctions are symmetric (i.e.\ with $p=1/2$) Krawtchouk polynomials.
For some plots of these discrete wavefunctions, we refer to~\cite{Atak2001,Atak2005}.
Note that these discrete wavefunctions tend to the continuous wavefunctions of the canonical quantum oscillator for large values of $j$. 
The eigenvalues of ${\hat p}$ are the same as for ${\hat q}$ and also given by $p_k=-j+k$. The corresponding coefficients in~\eqref{pk} are given by
\begin{equation}
\psi_n(p) = -\ii^{n+1} \phi_n(p).
\end{equation}

In order to compute the Wigner distribution function $W(n;p,q)$ (or the corresponding matrix ${\mathbf W}(n)$), one performs the following steps:
\begin{enumerate}
\item Using the known matrix representations of ${\hat p}$ and ${\hat q}$, compute the matrix of
${\hat G}_{a,b}(\hat p, \hat q)$ for all $a,b=0,1,\ldots,N$. 
\item Take the $(n,n)$-diagonal element of the matrix ${\hat G}_{a,b}(\hat p, \hat q)$ and put it in the matrix ${\mathbf Z}(n)$ on position $(a,b)$, according to~\eqref{defZ}.
\item Once the matrix ${\mathbf Z}(n)$ is determined, multiply by the inverses of the Vandermonde matrices
${\mathbf V}(-j,-j+1,\ldots,j)$ according to~\eqref{main}, and the result is ${\mathbf W}(n)$.
\end{enumerate}
In step~1, one can simplify the computations by taking powers $X^r$ of the matrix $X=\lambda{\hat p}+\mu{\hat q}$, and
then the appropriate coefficient of $\lambda^a\mu^b$ in $X^r$ ($r=a+b$) yields the desired matrix  
${\hat G}_{a,b}(\hat p, \hat q)$.

As a simple example, let $j=1$ (so $N=2$). Then 
\[
{\hat q} = \left( \begin{array}{ccc} 0 & \frac{1}{\sqrt{2}} & 0 \\ \frac{1}{\sqrt{2}} & 0 & \frac{1}{\sqrt{2}} \\ 0 & \frac{1}{\sqrt{2}} & 0 \end{array}\right),
\qquad
{\hat p} = \left( \begin{array}{ccc} 0 & -\frac{\ii}{\sqrt{2}} & 0 \\ \frac{\ii}{\sqrt{2}} & 0 & -\frac{\ii}{\sqrt{2}} \\ 0 & \frac{\ii}{\sqrt{2}} & 0 \end{array}\right).
\]
Thus we obtain for the matrices ${\hat G}_{a,b}(\hat p, \hat q)$:
\begin{align*}
& {\hat G}_{0,0}= \left( \begin{array}{ccc} 1 & 0 & 0 \\ 0 & 1 & 0 \\ 0 & 0 & 1 \end{array}\right), \qquad
{\hat G}_{1,0}=  \left( \begin{array}{ccc} 0 & -\frac{\ii}{\sqrt{2}} & 0 \\ \frac{\ii}{\sqrt{2}} & 0 & -\frac{\ii}{\sqrt{2}} \\ 0 & \frac{\ii}{\sqrt{2}} & 0 \end{array}\right), \qquad
{\hat G}_{0,1} = \left( \begin{array}{ccc} 0 & \frac{1}{\sqrt{2}} & 0 \\ \frac{1}{\sqrt{2}} & 0 & \frac{1}{\sqrt{2}} \\ 0 & \frac{1}{\sqrt{2}} & 0 \end{array}\right), \\
& {\hat G}_{2,0}= \left( \begin{array}{ccc} \frac12 & 0 & -\frac12 \\ 0 & 1 & 0 \\ -\frac12 & 0 & \frac12 \end{array}\right), \qquad
{\hat G}_{1,1}= \left( \begin{array}{ccc} 0 & 0 & -\frac{\ii}{2} \\ 0 & 0 & 0 \\ \frac{\ii}{2} & 0 & 0 \end{array}\right), \qquad
{\hat G}_{0,2}= \left( \begin{array}{ccc} \frac12 & 0 & \frac12 \\ 0 & 1 & 0 \\ \frac12 & 0 & \frac12 \end{array}\right), \\
& {\hat G}_{2,1} = \left( \begin{array}{ccc} 0 & \frac{1}{3\sqrt{2}} & 0 \\ \frac{1}{3\sqrt{2}} & 0 & \frac{1}{3\sqrt{2}} \\ 0 & \frac{1}{3\sqrt{2}} & 0 \end{array}\right), \qquad
{\hat G}_{1,2}=  \left( \begin{array}{ccc} 0 & -\frac{\ii}{3\sqrt{2}} & 0 \\ \frac{\ii}{3\sqrt{2}} & 0 & -\frac{\ii}{3\sqrt{2}} \\ 0 & \frac{\ii}{3\sqrt{2}} & 0 \end{array}\right), \qquad
{\hat G}_{2,2}= \left( \begin{array}{ccc} \frac16 & 0 & 0 \\ 0 & \frac13 & 0 \\ 0 & 0 & \frac16 \end{array}\right).
\end{align*}
So one finds, e.g.
\[
{\mathbf Z}(0)= \left( \begin{array}{ccc} 1 & 0 & \frac12 \\ 0 & 0 & 0 \\ \frac12 & 0 & \frac16 \end{array}\right), \qquad
{\mathbf Z}(1)= \left( \begin{array}{ccc} 1 & 0 & 1 \\ 0 & 0 & 0 \\ 1 & 0 & \frac13 \end{array}\right).
\]
Then, multiplying by the inverses of Vandermonde matrices yields:
\begin{equation}
{\mathbf W}(0) = \left( \begin{array}{ccc} \frac{1}{24} & \frac{1}{6} & \frac{1}{24} \\ \frac{1}{6} & \frac{1}{6} & \frac{1}{6} \\ \frac{1}{24} & \frac{1}{6} & \frac{1}{24} \end{array}\right), \qquad
{\mathbf W}(1) = \left( \begin{array}{ccc} \frac{1}{12} & \frac{1}{3} & \frac{1}{12} \\ \frac{1}{3} & -\frac{2}{3} & \frac{1}{3} \\ \frac{1}{12} & \frac{1}{3} & \frac{1}{12} \end{array}\right).
\end{equation}
The example illustrates the computation procedure, however the dimension of the representation is too small to observe some general properties, and in particular to compare to the Wigner distribution functions for the canonical oscillator.

For this purpose, we have performed the same computations for $j=12$.
The computation is similar to the $j=1$ example, just more involved. 
The result consists of $25\times 25$-matrices for ${\mathbf W}(n)$ ($n=0,1,2,\ldots$).
These matrices are of course too big to print here, but instead we give a matrix plot of them for $n=0,1,2$ in Figure~1.
Consider, for example, the top left plot of Figure~1, which is the matrix plot of ${\mathbf W}(0)$. 
The horizontal axes correspond to $p$ and $q$, both discrete variables taking values from $-12$ up to $+12$ in steps of~1. 
For each of these $25\times 25$ values in the horizontal plane, 
we plot the value of the Wigner function $W(n;p,q)$ in the vertical direction. 
Rather than just plotting points or dots, we plot these values as a `histogram' in order to visualize the shape.
Such a matrix plot is performed for ${\mathbf W}(0)$, ${\mathbf W}(1)$ and ${\mathbf W}(2)$ in the left column of Figure~1.

Since the $\mathfrak{su}(2)$ finite oscillator is often considered as a finite version of the canonical oscillator
(to which it tends when $j\rightarrow\infty$), it is interesting to compare the shapes of our discrete Wigner functions 
with the usual continuous Wigner function of the canonical oscillator, given by~\cite{Hillery1984}
\begin{equation}
W_n(p,q) = \frac{(-1)^n}{\pi} \ee^{-p^2-q^2}\; L_n(2p^2+2q^2),
\end{equation}
where $L_n$ is the Laguerre polynomial.
These canonical Wigner functions, and their plots, are well known.
For comparison with the discrete case, we have plotted the canonical Wigner functions for $n=0,1,2$ in the right column of Figure~1.
At first sight, the similarity of the discrete shapes and continuous shapes is striking.
Note that of course the scaling is different in the discrete and continuous plots.
In the continuous plots, we have depicted the function for $p$ and $q$ (in the horizontal plane) running from $-4$ to $4$.
Due to the fact that in the canonical case we have 
\[
\int\int W_n(p,q) dpdq = 1,
\]
and in the discrete case we have
\[
\sum_{p=-j}^{j} \sum_{q=-j}^j W(n;p,q) =1,
\]
the scaling is different and the ``heights'' of the plots in the left and right column of Figure~1 do not match.

Since the 3D plots in Figure~1 also ``hide'' a part of the shape, it is worth making also a density plot of these functions.
This has been performed in Figure~2. 
As in Figure~1, the left column shows the density plot for the discrete functions $W(n;p,q)$ ($n=0,1,2$), and the right column for
the continuous functions $W_n(p,q)$.
For the discrete density plots, the $p$ and $q$ values run again from $-12$ up to $+12$ in steps of~1. 
The gray values of the density plots are relative, in the sense that white color corresponds to the maximum value of $W(n;p,q)$, and black to the minimum value of $W(n;p,q)$.
For the continuous density plots, $p$ and $q$ run from $-4$ to $4$, and the function value is again relative according to the same convention.

Note that in the current example, the spectrum of ${\hat p}$ and ${\hat q}$ is symmetric (satisfying~\eqref{sym}), 
and that the matrix representations of ${\hat p}$ and ${\hat q}$ satisfy~\eqref{tridia}.
So the conditions of Proposition~3 are satisfied, and hence the Wigner function satisfies the phase-space symmetries~\eqref{Wsym}.
These symmetries are also clear in the plots.

In the case of the canonical quantum oscillator, the Wigner functions assume also negative values, so they are only ``quasiprobability''
distributions. Only for the ground state, $W_0(p,q)$ assumes positive values only.
In the discrete case, even for $n=0$ the values of $W(n;p,q)$ can already be negative. For example, for $j=5/2$, $W(0;j,j)<0$.
The absolute values of these negative numbers near the corners of the square grid are however so small that they are not or hardly
visible in the plots of Figures~1 and~2.

\section{Conclusions}

We have developed in this paper a new definition of a Wigner function for a finite quantum system, i.e.\ 
a quantum system in which the position operator ${\hat q}$ has a finite multiplicity-free spectrum $\{q_0,q_1,\ldots,q_N\}$ and
the momentum operator ${\hat p}$ also has a finite multiplicity-free spectrum $\{p_0,p_1,\ldots,p_N\}$.
When the system is in a stationary state $|n\rangle$ (eigenstate of some Hamiltonian operator ${\hat H}$; $n=0,1,\ldots,N$),
the corresponding Wigner function $W(n;p,q)$ is defined on a square grid $(p,q)\in \{p_0,p_1,\ldots,p_N\} \times
\{q_0,q_1,\ldots,q_N\}$, in other words is can be considered as an $(N+1)\times (N+1)$ matrix.
Our definition is based on a natural assumption, \eqref{Wab}, requiring that the distribution averages for all
functions $p^aq^b$ ($a,b=0,\ldots,N$) coincide with their quantum state averages for the corresponding operator form
(following Weyl's association).
This approach leads to a method by which the Wigner functions (or corresponding matrices) can be computed,
involving in particular Vandermonde matrices and their inverses.
Moreover, the Wigner functions satisfy a number of properties similar to those of continuous Wigner distribution
functions (listed, e.g.\ in~\cite{Hillery1984}).

In order to illustrate the practical computation of the discrete Wigner function, we have performed this 
for the $\mathfrak{su}(2)$ finite oscillator model. 
This is a popular finite oscillator for which all relevant operators (Hamiltonian, position, momentum) act in 
a space of dimension $2j+1$. 
The position and momentum operators have a simple spectrum $-j,-j+1,\ldots,+j$, and the corresponding discrete wavefunctions
are well known.
Moreover, for $j\rightarrow\infty$, these wavefunctions tend to the common continuous wavefunctions of the harmonic oscillator
in terms of Hermite functions.

We have computed the discrete Wigner function (for some value $j=12$), and plotted these values as a matrix plot over
the square grid formed by the discrete phase-space. 
These discrete plots are quite appealing, and their shape is compared to that of the continuous Wigner functions for
the canonical oscillator.

In this context, it should be mentioned that also in~\cite{Atak1998}, a Wigner operator and distribution function was
defined for the $\mathfrak{su}(2)$ model. The definition in~\cite{Atak1998} is certainly fundamental, and follows a group
theoretical approach, by means of an integration over the group $SU(2)$. Clearly, this approach works in case there is
an underlying Lie group. 
Our approach on the other hand, which may be less fundamental than the group theoretical approach, 
is certainly more general since there are no requirements for the quantum system relating it to a Lie group.

In order to understand and study our Wigner function approach better, it would be interesting to compute, for an
arbitrary $j$-value, the values $W(n;p_k,q_l)$ explicitly (say, as a function of $j$, $n$, $k$ and $l$).
This seems to be doable, but certainly harder than it looks at first sight (and probably involving
multiple hypergeometric series manipulations).
We hope to report on such computations in the near future.

\newpage
\begin{figure}[th]
\[
\begin{tabular}{cc}
\includegraphics[height=85mm,width=85mm]{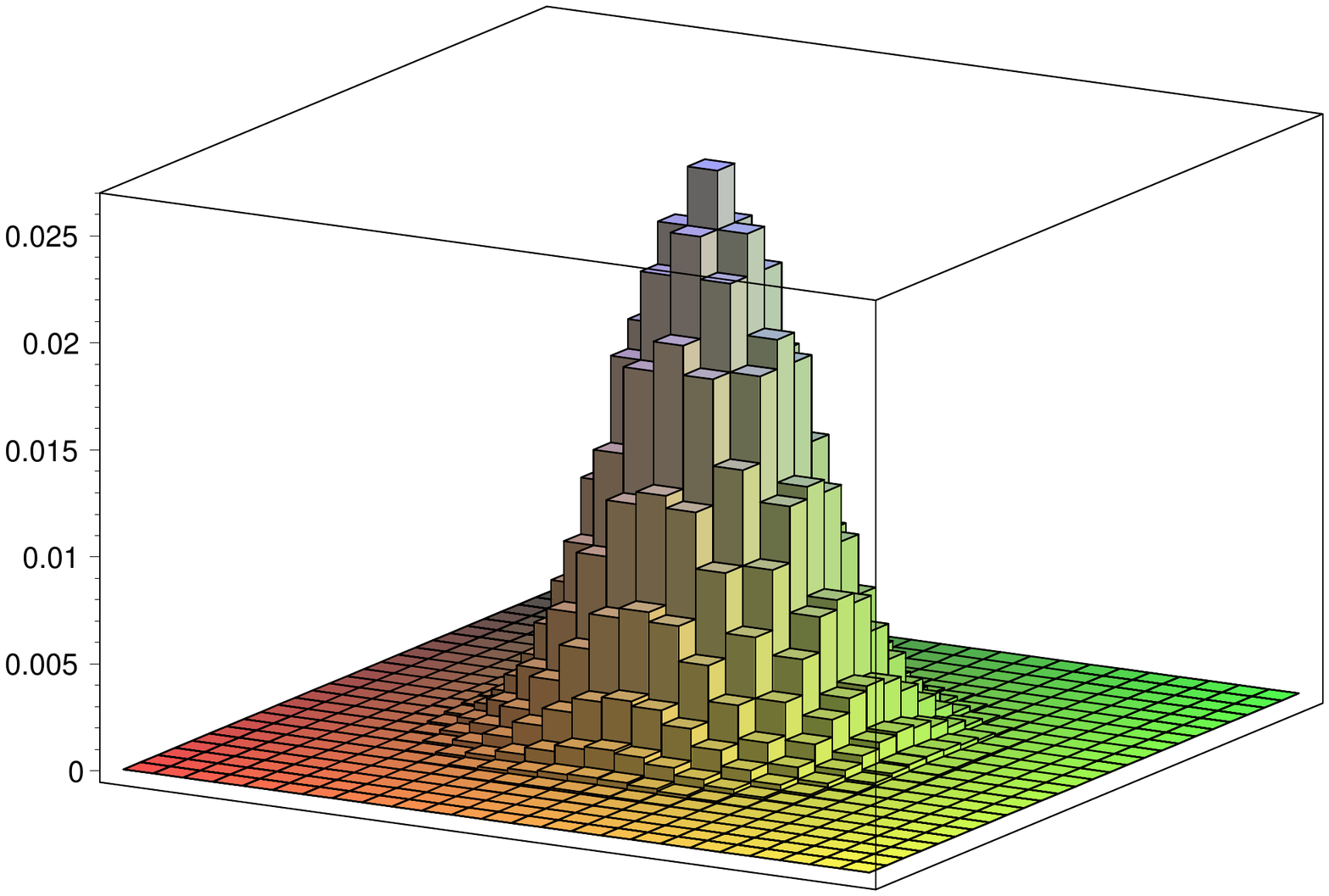}\hskip -10mm & \includegraphics[height=85mm,width=85mm]{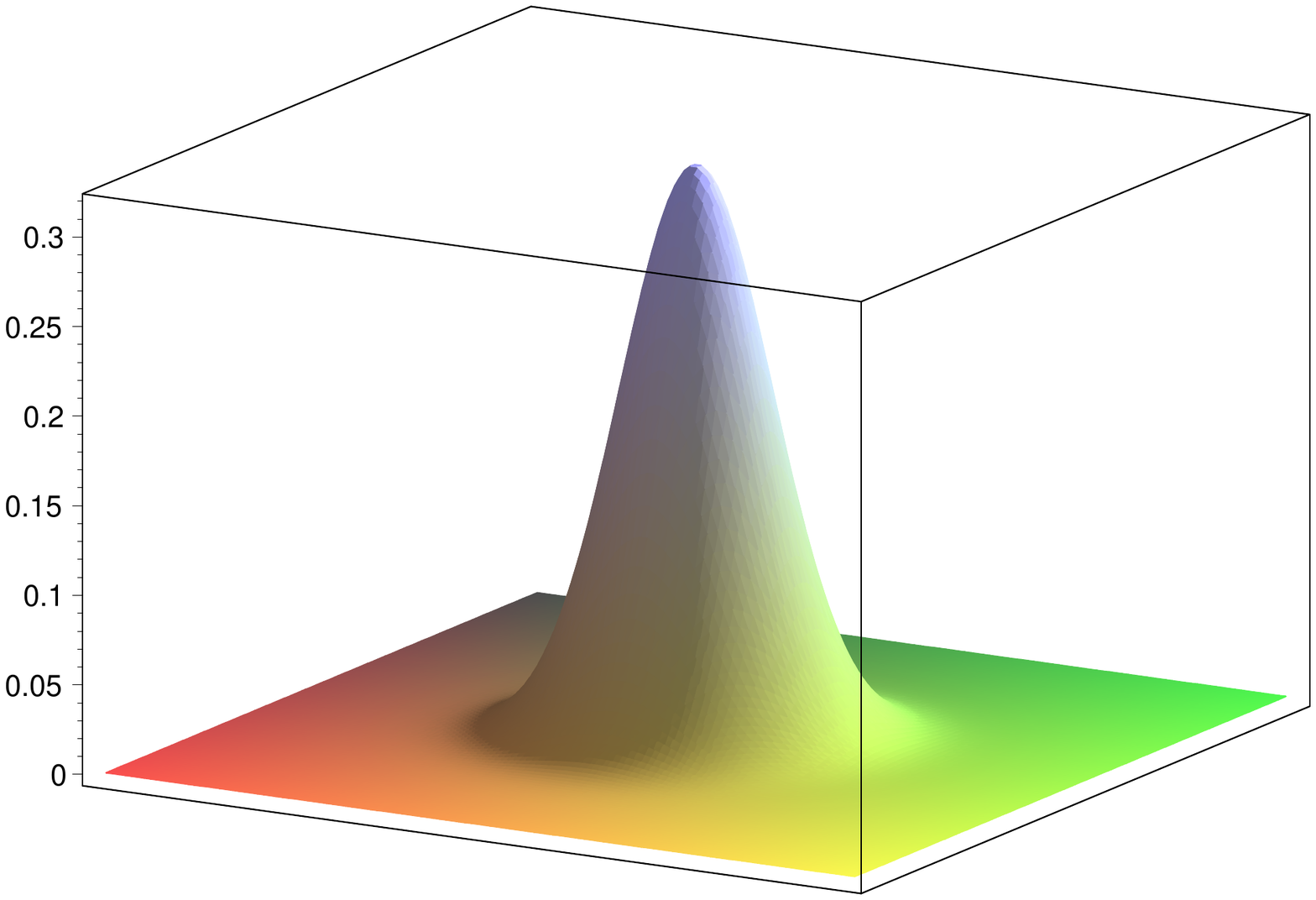}  \\[-20mm]
\includegraphics[height=85mm,width=85mm]{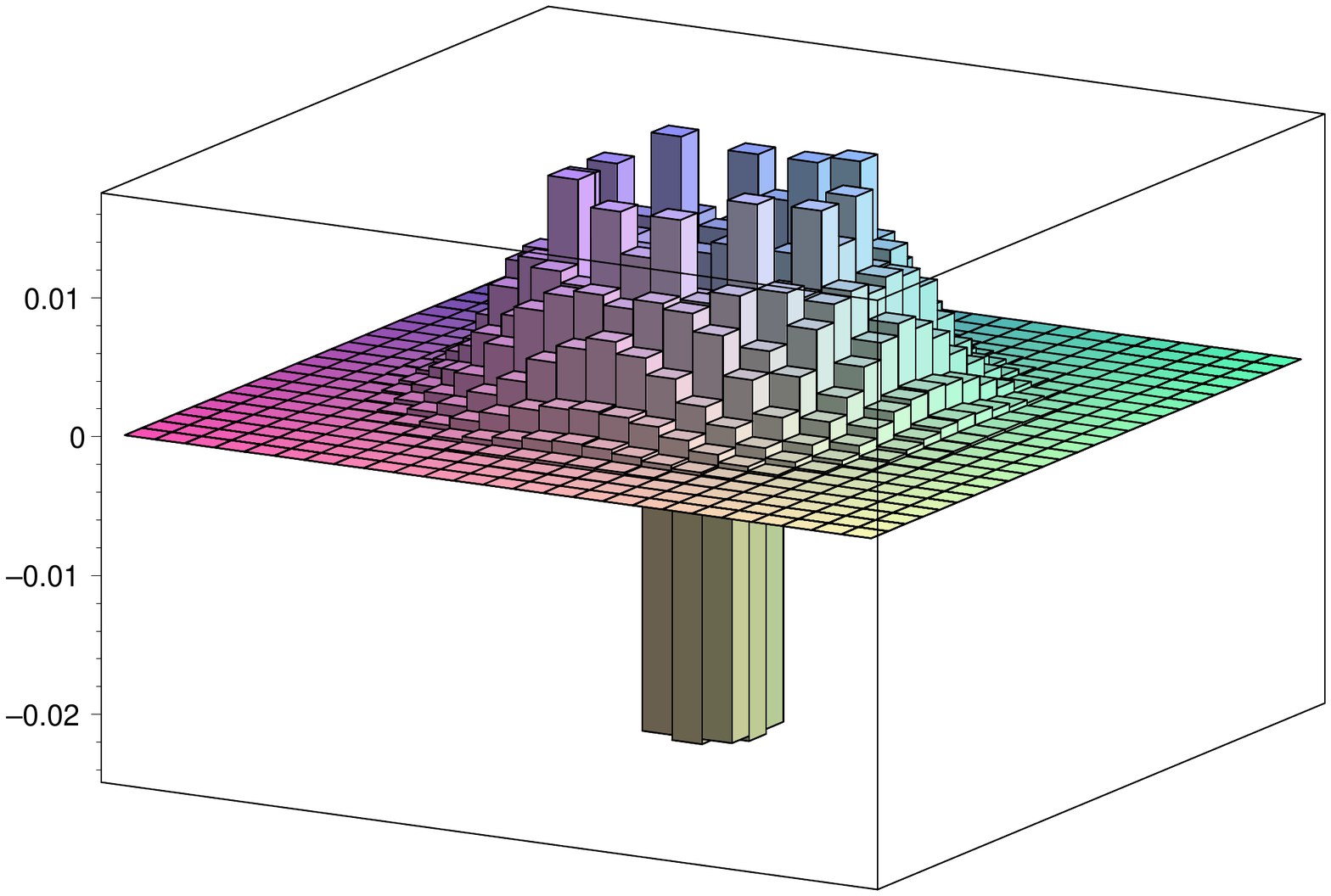}\hskip -10mm & \includegraphics[height=85mm,width=85mm]{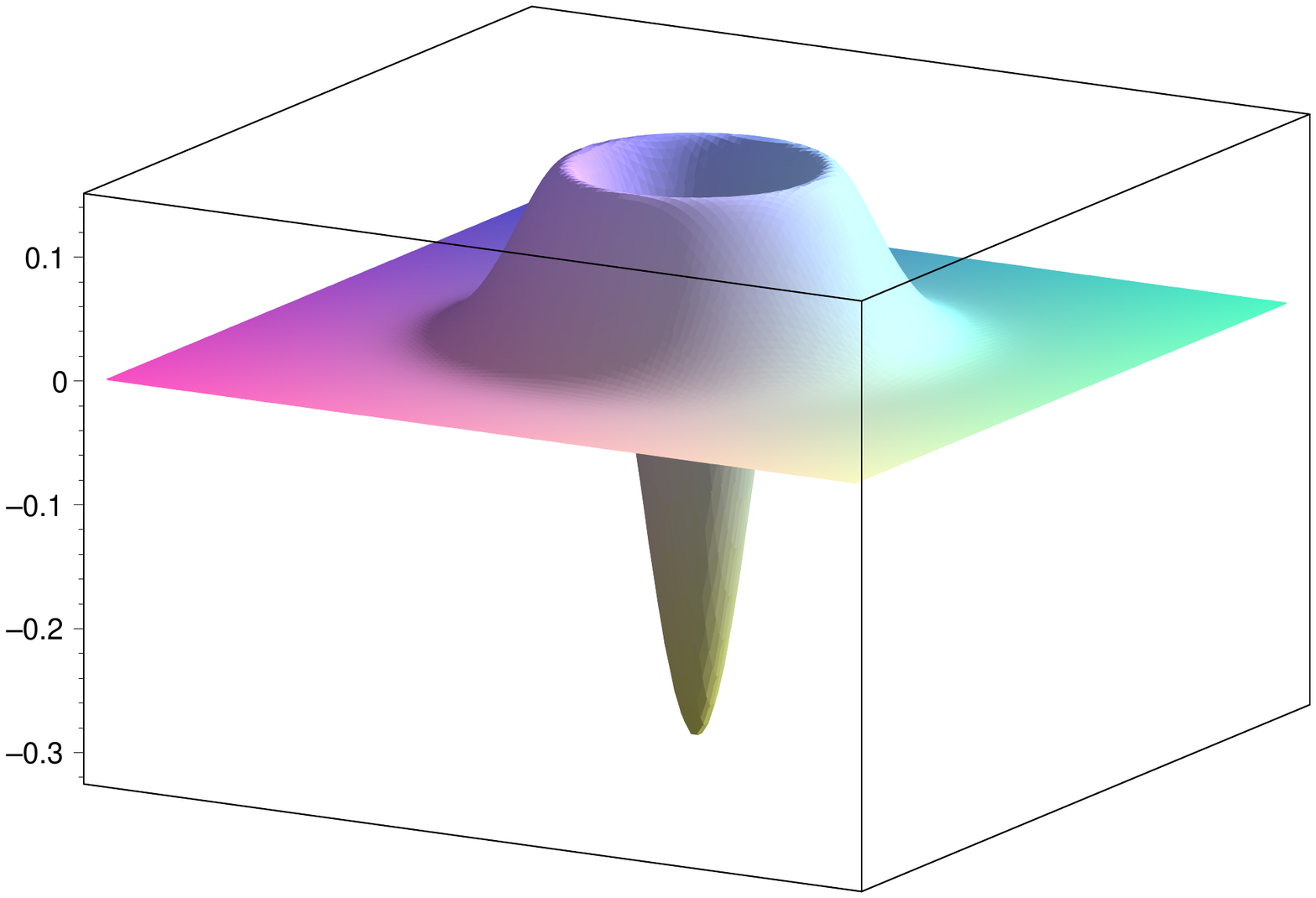}  \\[-20mm]
\includegraphics[height=85mm,width=85mm]{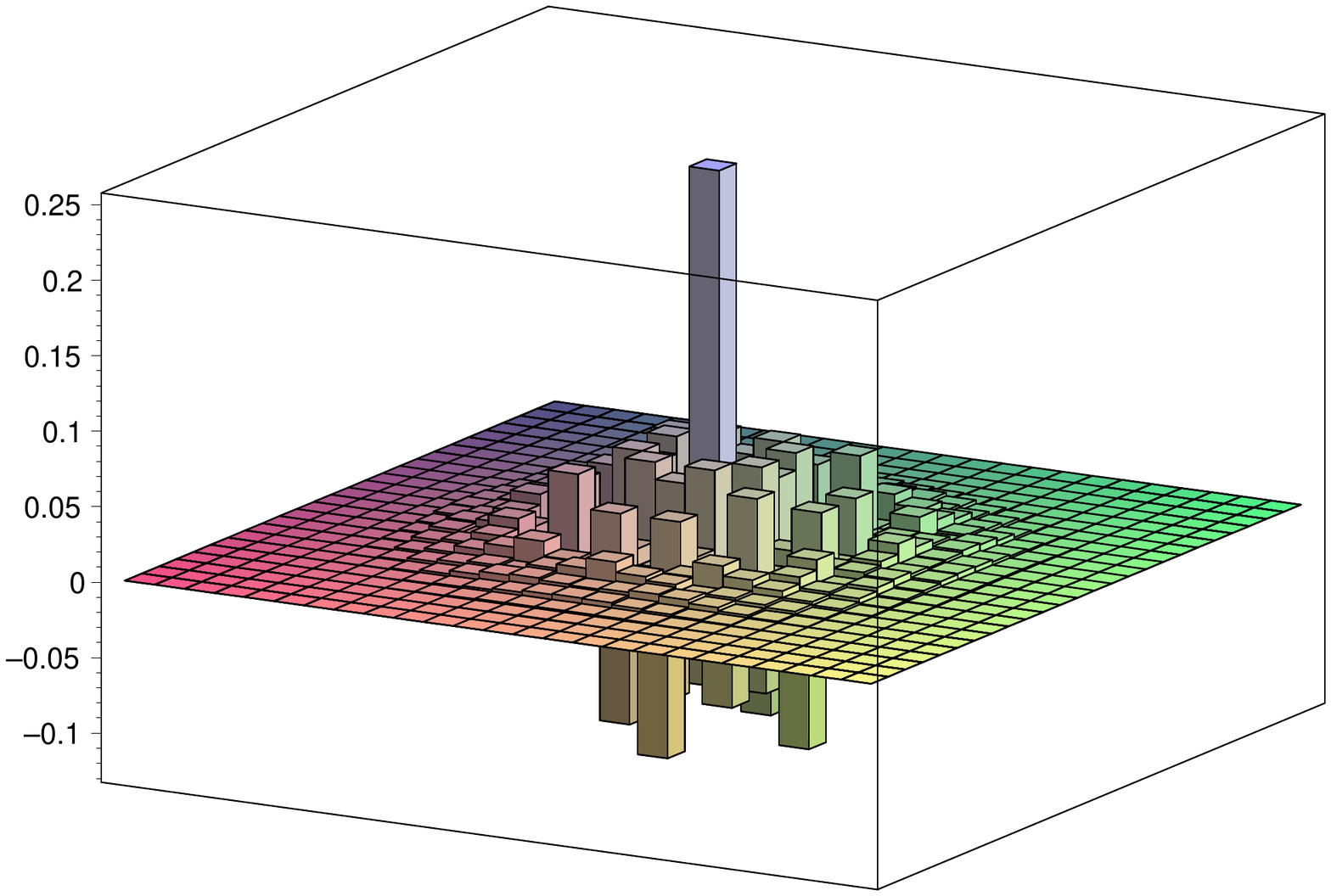}\hskip -10mm & \includegraphics[height=85mm,width=85mm]{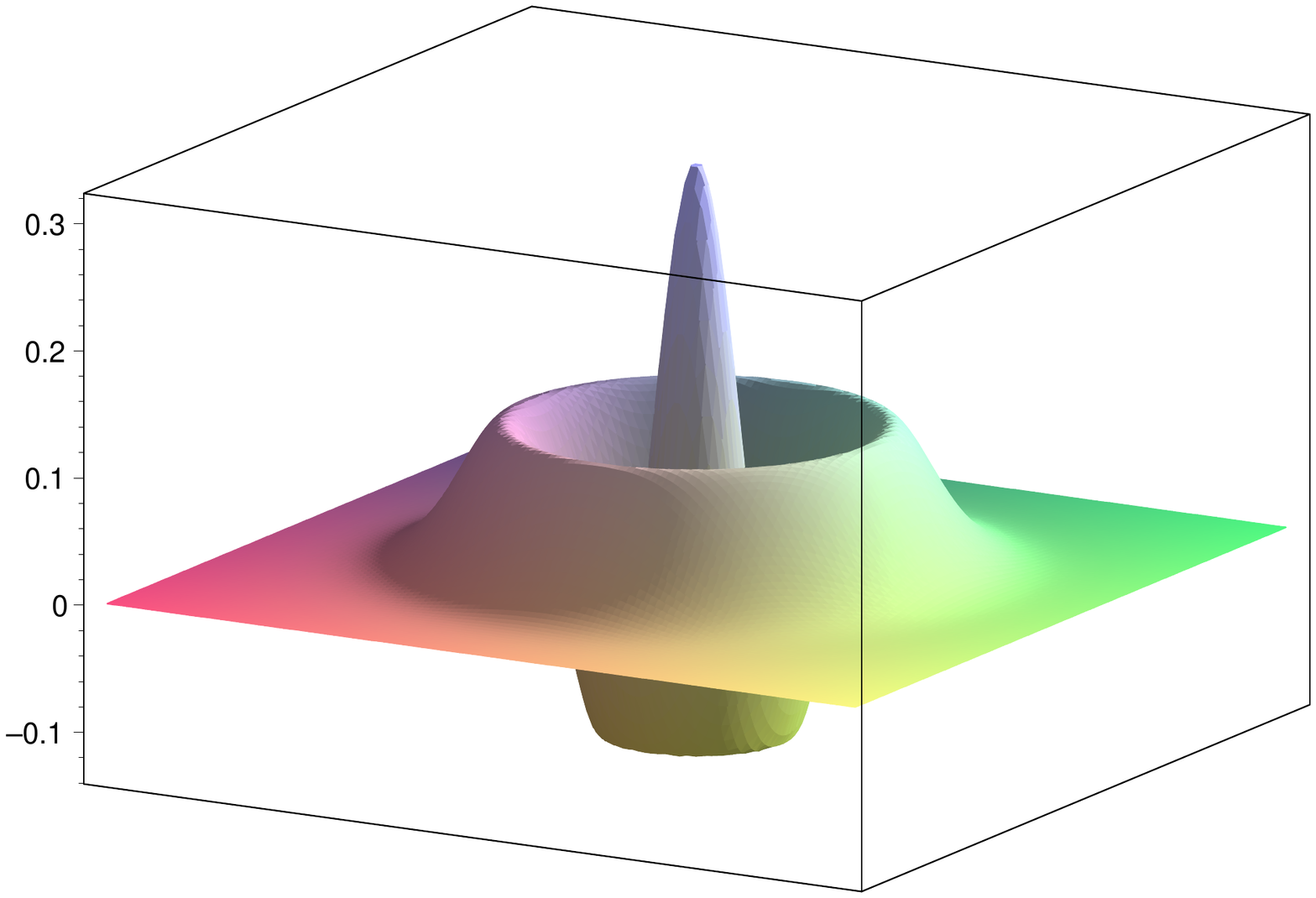}  \\[-20mm]
\end{tabular} 
\]
\caption{In the left column: matrix plots of the discrete Wigner functions $W(n;p,q)$ for $n=0$, $1$ and $2$ (all for $j=12$).
The discrete $p$ and $q$ values in the horizontal planes run from $-j$ up to $+j$.
In the right column: 3D surface plots of the canonical Wigner functions $W_n(p,q)$ for $n=0$, $1$ and $2$,
with $p$ and $q$ taking continuous values in the horizontal planes running from $-4$ up to $4$.}
\label{fig1}
\end{figure}

\newpage
\begin{figure}[th]
\[
\begin{tabular}{cc}
 & \\[-70mm]
\hskip -40mm \includegraphics[scale=0.5]{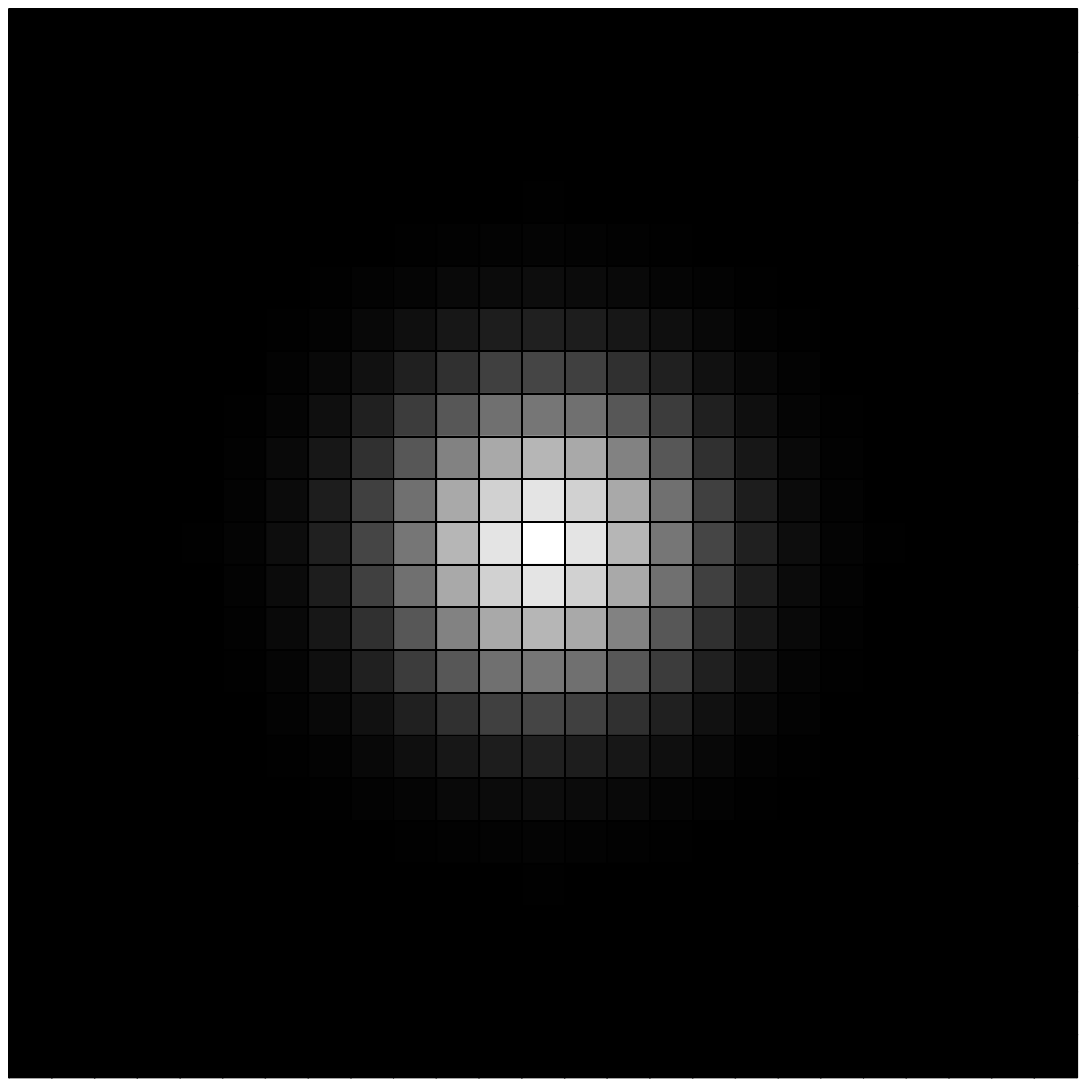} & \hskip -40mm \includegraphics[scale=0.5]{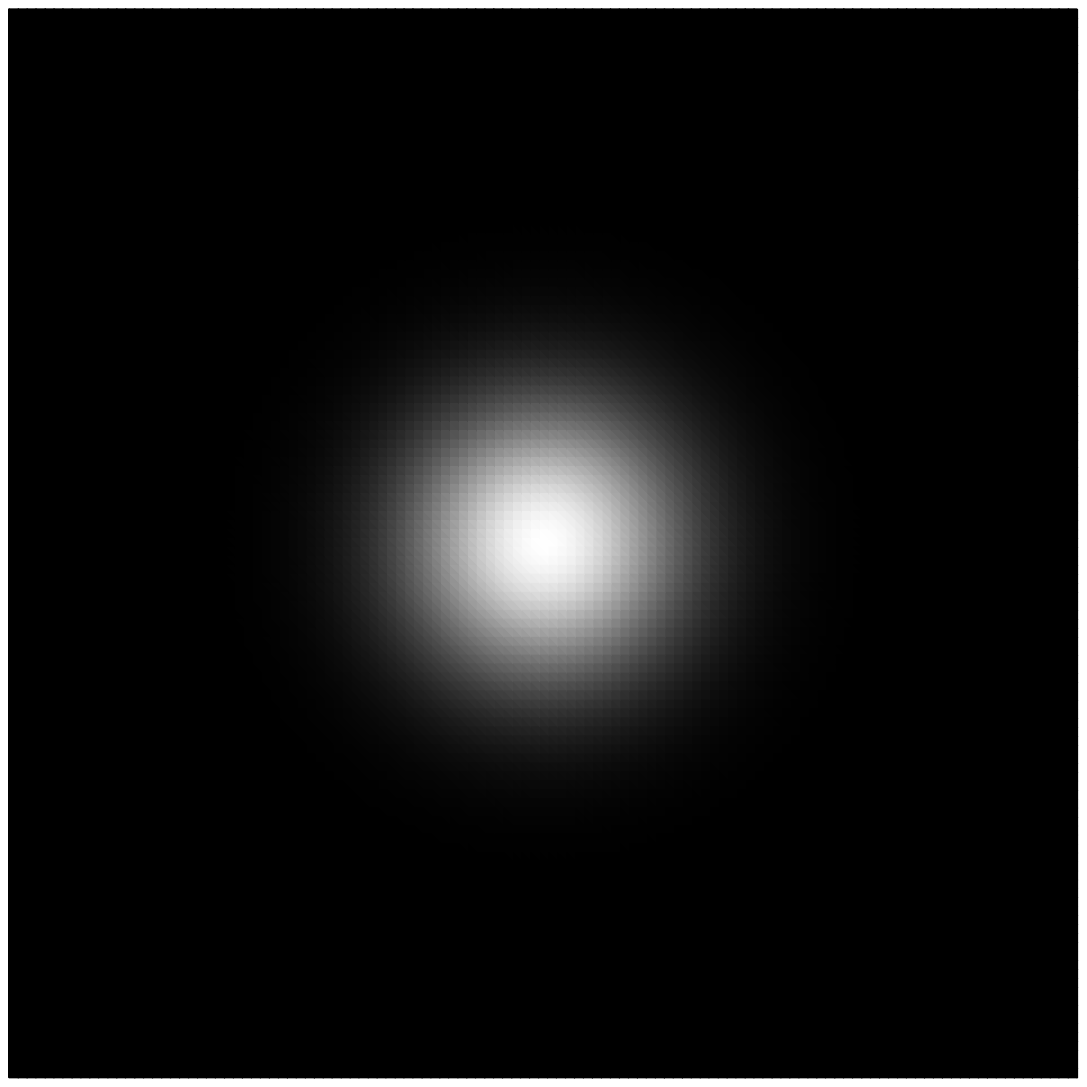}  \\[-80mm]
\hskip -40mm \includegraphics[scale=0.5]{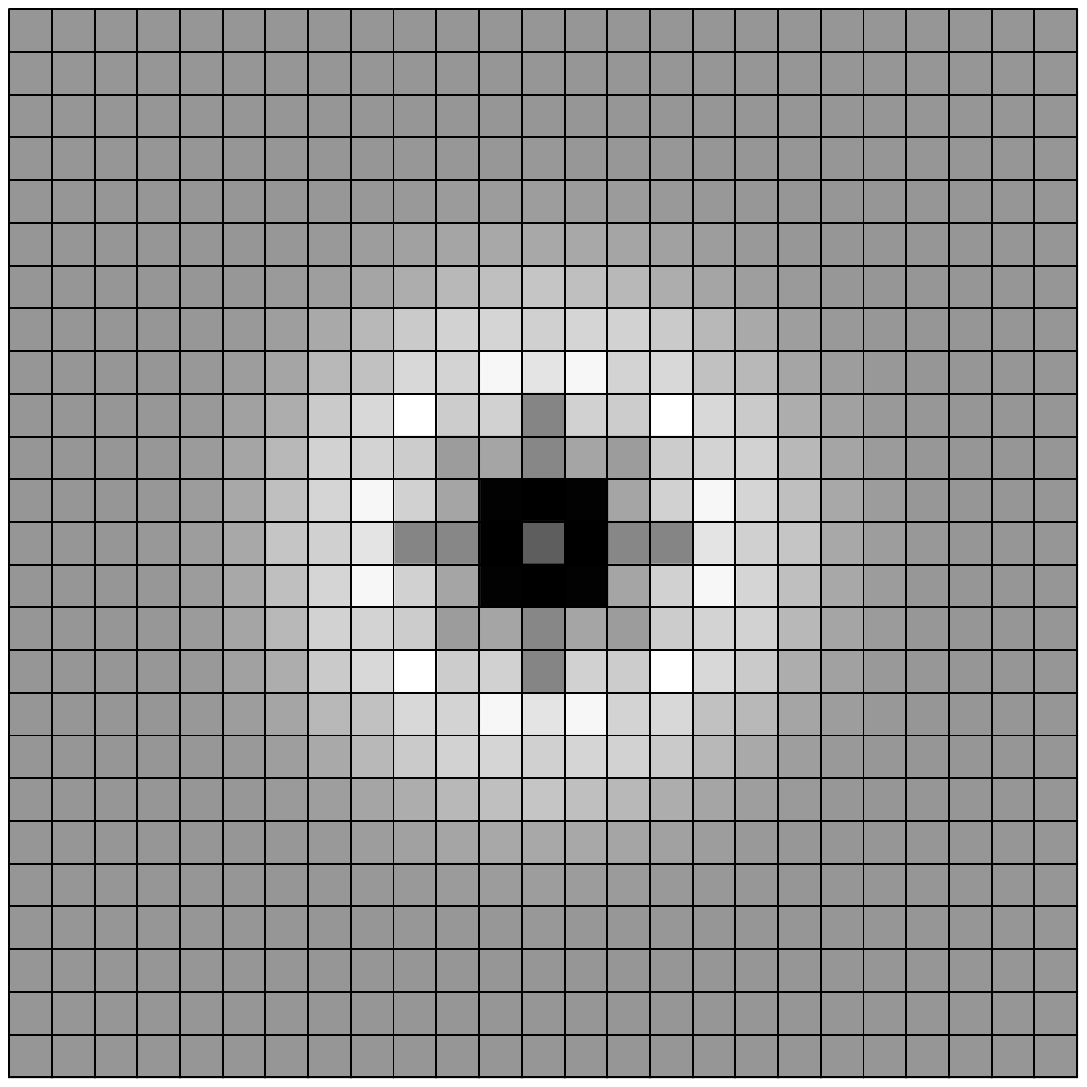} & \hskip -40mm \includegraphics[scale=0.5]{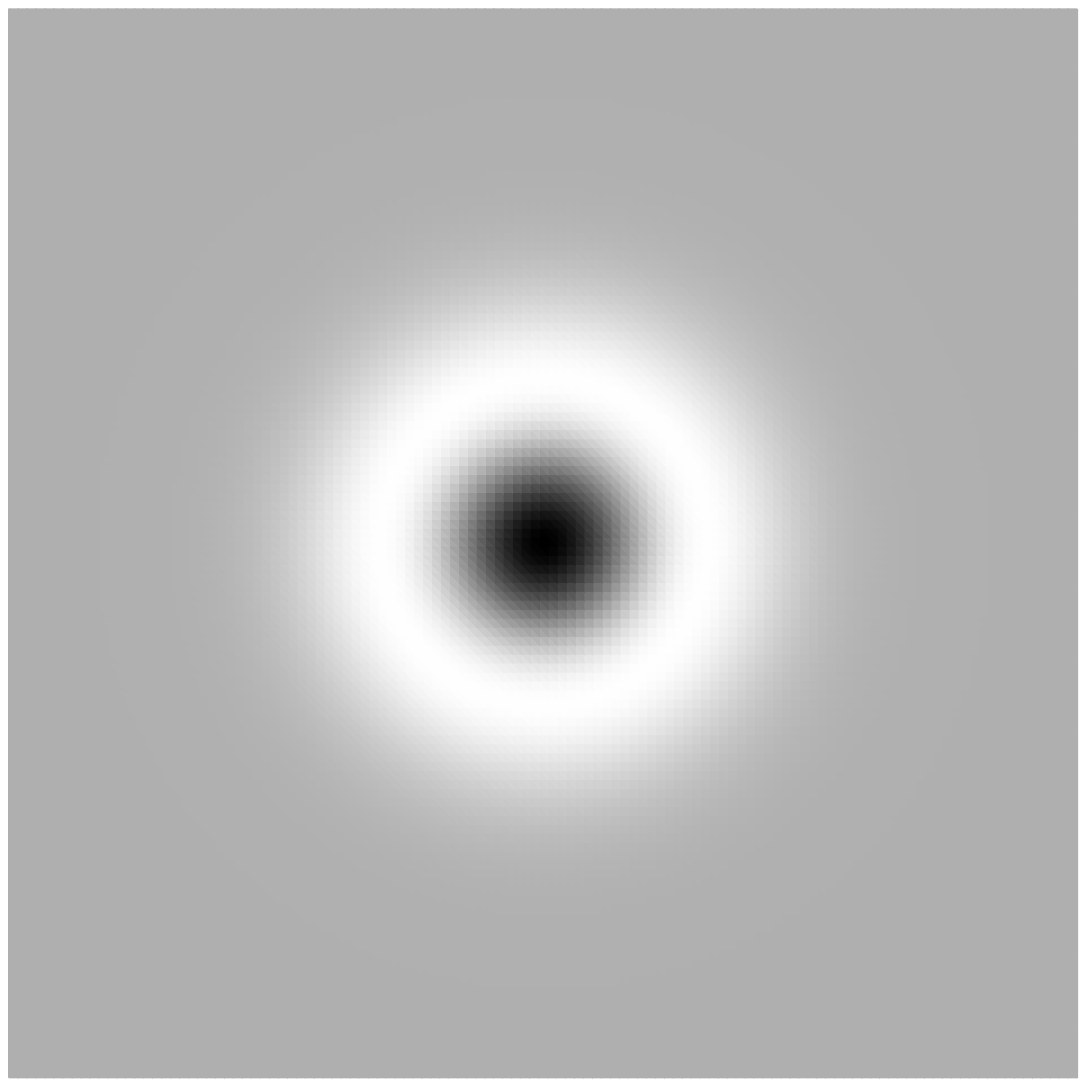}  \\[-80mm]
\hskip -40mm \includegraphics[scale=0.5]{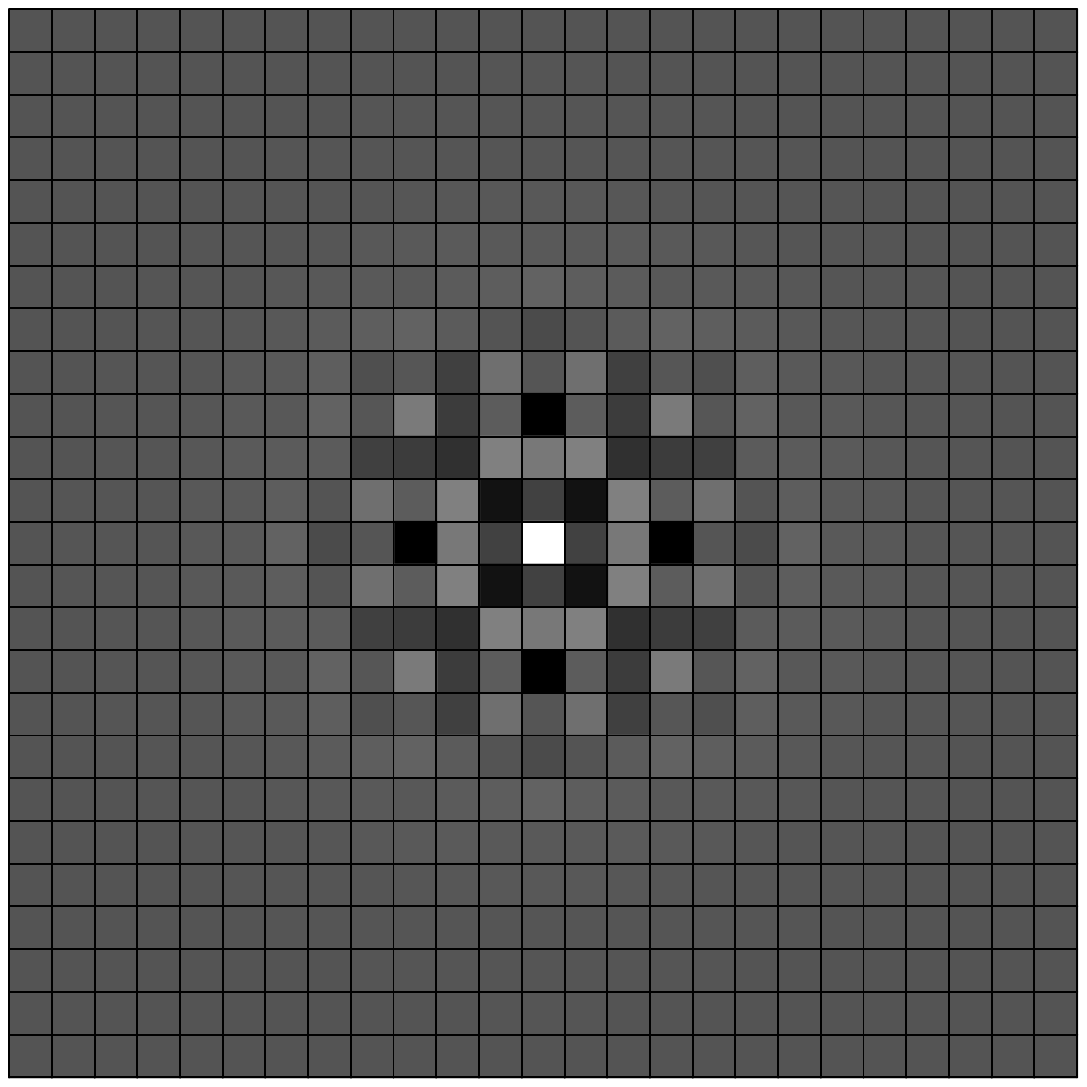} & \hskip -40mm \includegraphics[scale=0.5]{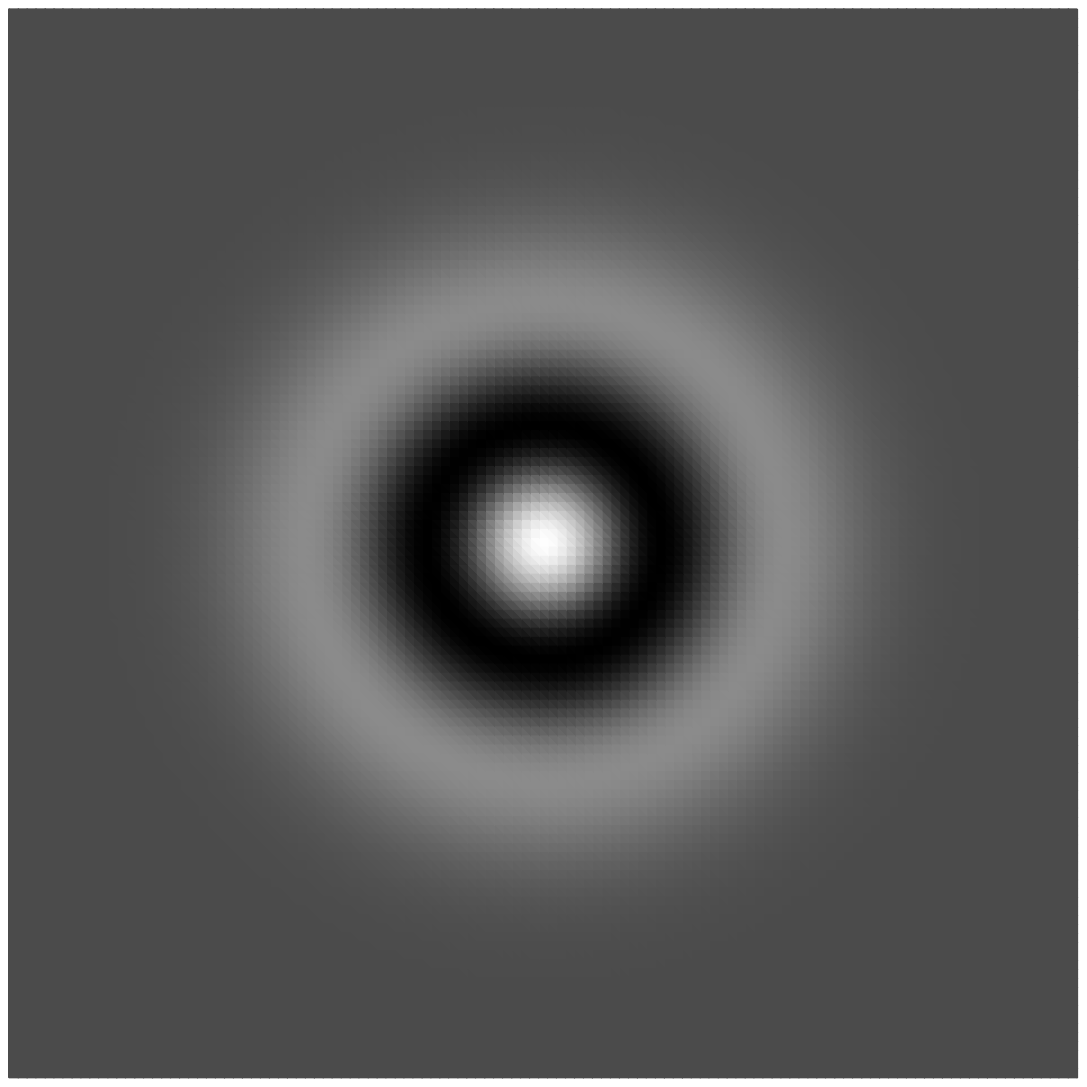}  \\
\end{tabular} 
\]
\caption{In the left column: density plots of the discrete Wigner functions $W(n;p,q)$ for $n=0$, $1$ and $2$ (all for $j=12$).
In the right column: density plots of the canonical Wigner functions $W_n(p,q)$ for $n=0$, $1$ and $2$.
The gray levels are relative in each plot: white corresponds to the maximum value of the function, black to the minimum value.}
\label{fig2}
\end{figure}

\end{document}